\documentclass{article}

 \usepackage{graphicx}

\usepackage{amssymb}
\usepackage{amsthm}
\usepackage{amsmath}

\def\beq{\begin{equation}}
\def\eeq{\end{equation}}
\def\bea{\begin{eqnarray}}
\def\eea{\end{eqnarray}}

\title{Transverse Spin and Transverse Momentum Effects at COMPASS}

 \author{Christian Schill\\
 \small on behalf of the COMPASS collaboration\\
 \small Physikalisches Institut der
\small Albert-Ludwigs-Universit\"at Freiburg\\
\small Hermann-Herder-Str. 3, 79104 Freiburg, Germany}

\date{January 2011}

\begin{document}
\maketitle

\begin{abstract}  The investigation of transverse spin and transverse momentum
effects in deep inelastic scattering is one of the key physics programs of the
COMPASS collaboration. In the years $2002-2004$ COMPASS took data scattering 
$160$~GeV muons on a transversely polarized $^6$LiD target. In $2007$, a 
transversely polarized NH$_3$ target was used. Three different channels to access the
transversity distribution function have been analyzed: The azimuthal
distribution of single hadrons, involving the Collins fragmentation function,
the azimuthal dependence of the plane containing hadron pairs, involving the
two-hadron interference fragmentation function, and the measurement of the
transverse polarization of lambda hyperons in the final state. Transverse quark
momentum effects in a transversely polarized nucleon have been investigated by
measuring the Sivers distribution function. Azimuthal asymmetries in
unpolarized semi-inclusive deep-inelastic scattering give important information
on the inner structure of the nucleon as well, and can be used to estimate both
the quark transverse momentum in an unpolarized nucleon and to access the
so-far unmeasured Boer-Mulders function. COMPASS has measured these asymmetries
in 2004 using spin-averaged $^6$LiD data. \end{abstract}

\section{Introduction}

The study of transverse polarization and transverse momentum dependent distribution
functions was initiated by Ralston and Soper in their study of Drell-Yan processes
\cite{Ralston}. In order to understand these effects in a QCD framework, the description of
the partonic structure of the nucleon has been extended to include the quark transverse spin
and transverse momentum $k_T$ \cite{Mulders, Aram1, Aram2}.  Recent data on single spin asymmetries in
semi-inclusive deep-inelastic scattering (SIDIS) off transversely polarized nucleon targets
\cite{COMPASS2,COMPASS,HERMES} triggered a lot of interest towards the transverse momentum dependent
and spin dependent distribution and fragmentation functions \cite{Efremov1, Efremov2,
Bacchetta, Aram,  Ma, AnselminoS}. 

The SIDIS cross-section in the one-photon exchange approximation contains eight
transverse-momentum dependent distribution functions \cite{Mulders}. Some of
these  can be extracted in SIDIS measuring the azimuthal distribution of the
hadrons in the final state \cite{Collins}. Three distribution functions  survive
upon integration over the transverse momenta: These are the quark momentum
distribution $q(x)$, the helicity distribution $\Delta q(x)$, and the
transversity distribution $\Delta_T q(x)$. The latter is defined
as the difference in the number density of quarks with momentum fraction $x$
with their transverse spin parallel to the nucleon spin and their transverse
spin anti-parallel to the nucleon spin \cite{Artru}. 

To access transversity in SIDIS, one has to measure the quark polarization,
i.e. use the so-called 'quark polarimetry'. Three complementary approaches  are
used at COMPASS:  a measurement of the single-spin asymmetries (SSA) in the
azimuthal distribution of the final state hadrons (the Collins asymmetry), a
measurement of the SSA in the azimuthal distribution of the plane containing 
final state hadron pairs (the two-hadron asymmetry), and a measurement of the
polarization of final state hyperons (the $\Lambda$-polarimetry). 

The chiral-odd  Boer-Mulders function is of special interest among the other
transverse-momentum dependent distribution functions \cite{Boer}. It describes
the transverse parton polarization in an unpolarized hadron. The Boer-Mulders
function generates azimuthal asymmetries in unpolarized SIDIS, together with
the so-called Cahn effect \cite{Cahn}, which arises from the fact that the
kinematics is non-collinear when $k_T$ is taken into account.

\section{The COMPASS experiment}

COMPASS is a fixed target experiment at the CERN SPS accelerator with a wide physics program
focused on the nucleon spin structure and on hadron spectroscopy. COMPASS investigates
transversity and the transverse momentum structure of the nucleon in semi-inclusive
deep-inelastic scattering. A $160$~GeV muon beam is scattered off a transversely polarized
NH$_3$ or $^6$LiD  target. The scattered muon and the produced hadrons are detected in a
50~m long wide-acceptance forward spectrometer with excellent particle identification
capabilities \cite{Experiment}. A variety of tracking detectors is used to cope with the different
requirements of position accuracy and rate capability at different angles. Particle
identification is provided by a large acceptance RICH detector, two  electromagnetic and
hadronic calorimeters, and muon filters. 

The polarized $^6$LiD target is split into two cylindrical cells along the beam
direction. The two cells are polarized in opposite direction. The
polarized NH$_3$ target consists of three cells (upstream, central and
downstream) of 30, 60 and 30 cm length, respectively. The upstream and
downstream cell are polarized in one direction while the middle cell is
polarized oppositely. 

The polarization of the $^6$LiD target is $48\%\pm 5\%$
and of the NH$_3$ target about  $90$\%. The amount of polarized material in the
target, the so called dilution factor, of the $^6$LiD target is $0.38$. The
dilution factor of the ammonia target is $0.15$ in bins of the hadron
fractional energy $z$ and the hadron transverse momentum $p_T$, while it
increases with Bjorken $x$ from $0.14$ to $0.17$. The direction of the target
polarization was reversed every five days. The asymmetries are analyzed using
at the same time data from two time periods with opposite polarization and 
from the different target cells. The data have been selected requiring a good
stability of the  spectrometer within one and between consecutive periods of
data taking. 

To select DIS events, kinematic cuts of the squared four momentum transfer
$Q^2>1$~(GeV/c)$^2$, the hadronic invariant mass $W>5$~GeV/c$^2$ and the fractional energy
transfer of the muon $0.1<y<0.9$ were applied.

\section{The Collins asymmetry}

\begin{figure}
\hspace*{-4mm}\includegraphics[width=1.02\textwidth]{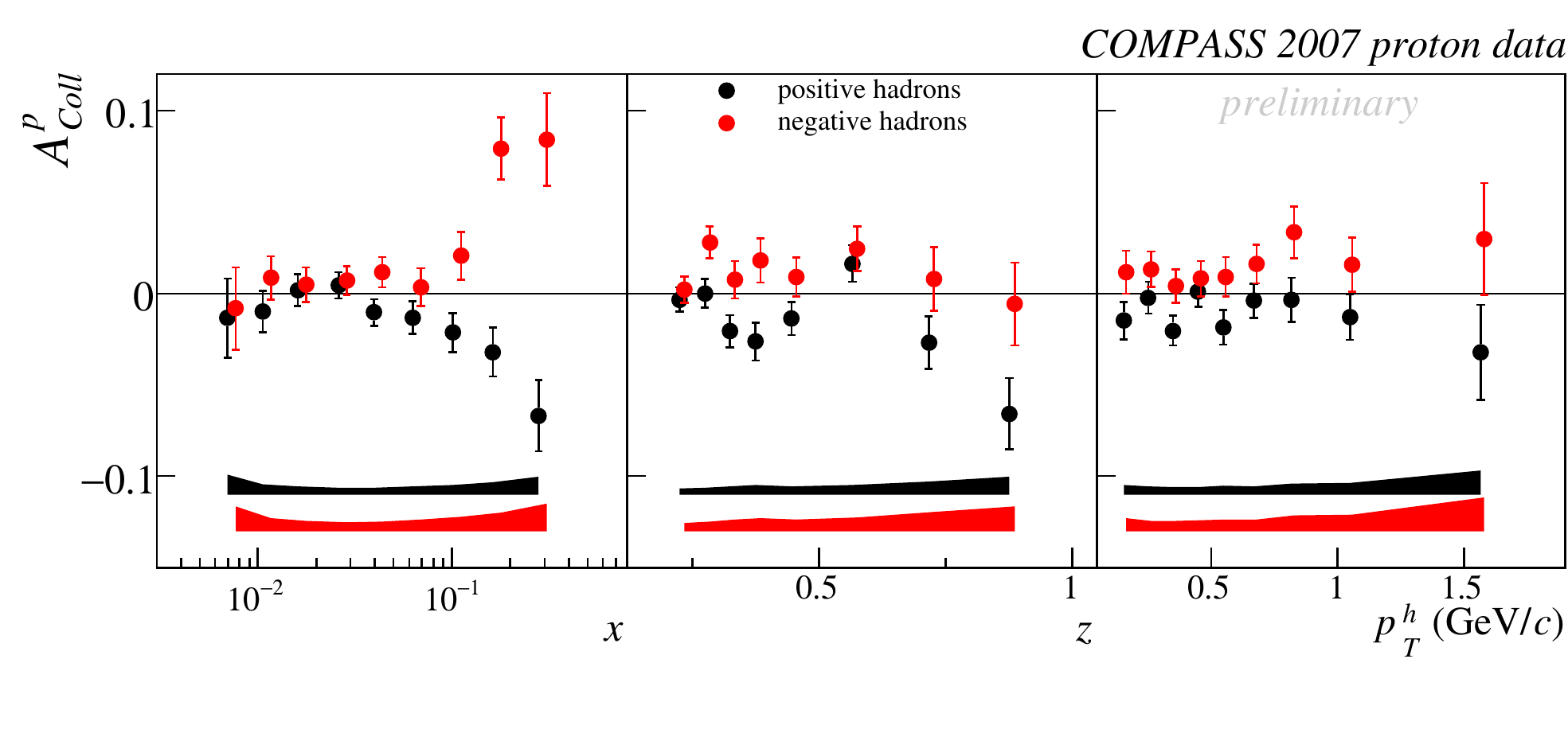}
\hspace*{-4mm}\includegraphics[width=1.02\textwidth]{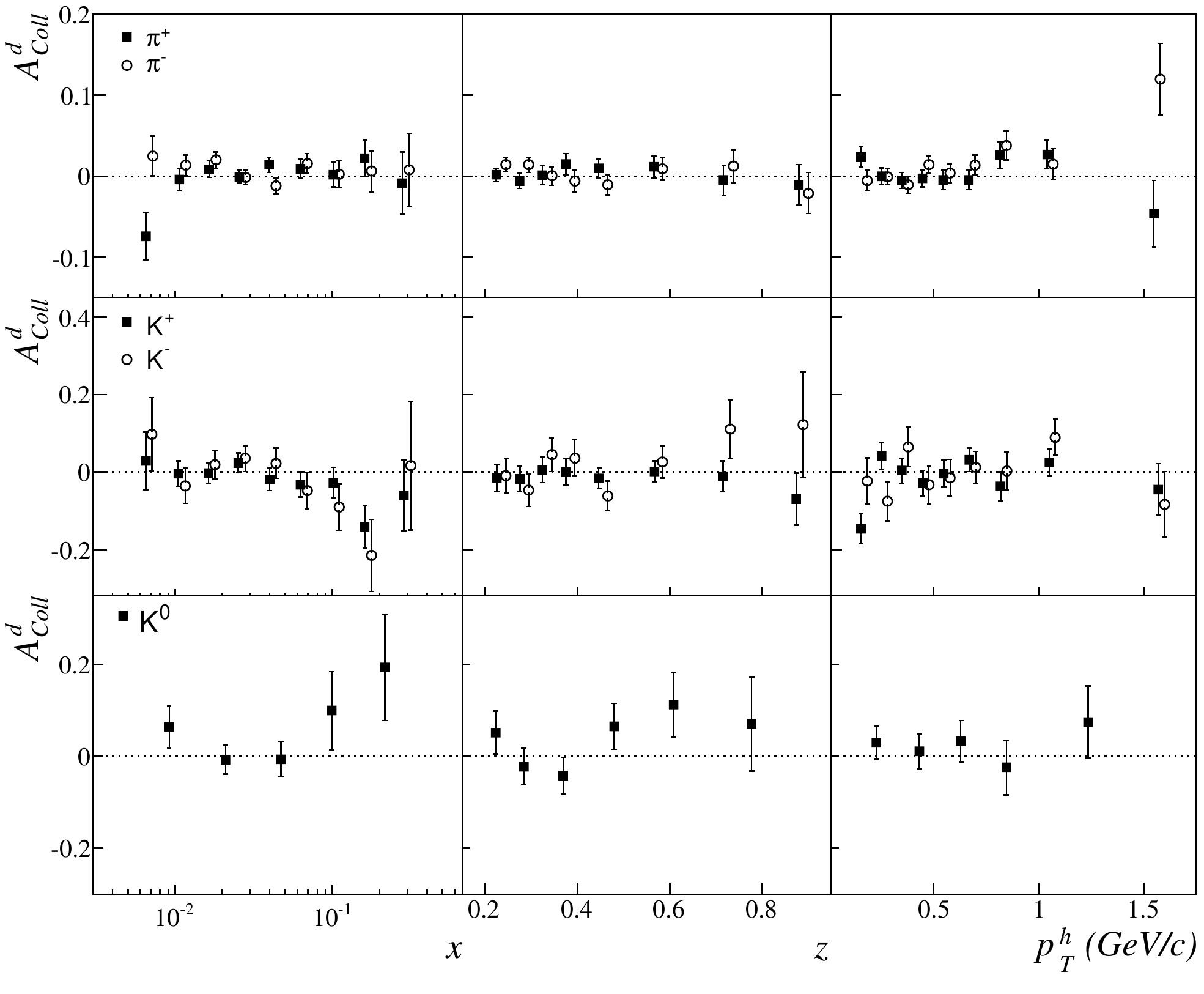}
\caption{{\it Upper panel:} Collins asymmetry on the proton for unidentified positive (black) and negative
(red) hadrons as a function of $x$, $z$, and $p_T$ as published in \cite{COMPASS2}. The bands indicate the
systematic uncertainty of the measurement.
{\it Lower panel:} Collins asymmetry on the deuteron for  positive (filled squares) and negative
(open circles) pions and kaons and $K^0$ as a function of $x$, $z$, and $p_T$ as published in \cite{COMPASS}.}
\label{Collins}
\end{figure}

In semi-inclusive deep-inelastic scattering the transversity
distribution $\Delta_Tq(x)$ can be measured in combination with the
chiral odd Collins fragmentation function
$\Delta^0_TD_q^h(x)$. According to Collins, the
fragmentation of a transversely polarized quark into an unpolarized
hadron generates an azimuthal modulation of the hadron distribution 
with respect to the lepton scattering plane \cite{Collins}. The hadron
yield $N(\Phi_{Coll})$ can be written as:
\begin{equation}
N(\Phi_{Coll})=N_0\cdot (1+f\cdot P_t\cdot D_{NN}\cdot A_{Coll}\cdot \sin\Phi_{Coll}),
\label{equ:Collins}
\end{equation}
where $N_0$ is the average hadron yield, $f$ the fraction of
polarized material in the target, $P_t$ the target polarization,   $A_{Coll}$
the Collins asymmetry, $D_{NN}=(1-y)/(1-y+y^2/2)$ the depolarization factor, and $y$ the fractional 
energy transfer of the muon. The angle $\Phi_{Coll}$ is the so called 
Collins angle. It is defined as  $\Phi_{Coll}=\phi_h+\phi_s-\pi$, the sum of the
hadron azimuthal angle $\phi_h$ and the target spin vector azimuthal angle
$\phi_S$, both with respect to the lepton scattering plane \cite{Artru}.
 The measured Collins asymmetry $A_{Coll}$ can be factorized into a convolution of the
transversity distribution $\Delta_Tq(x)$ and the
Collins fragmentation function $\Delta_T^0D_q^h(z, p_T)$, summed over all quark
flavors $q$:
\begin{equation}
A_{Coll}=\frac{\sum_q\,  e_q^2\cdot \Delta _Tq(x)\cdot \Delta_T^0D_q^h(z, p_T)}
{\sum_q\, e_q^2 \cdot q(x)\cdot D_q^h(z, p_T)}.
\end{equation}
Here, $e_q$ is the quark charge, $D^h_q(z, p_T)$ the unpolarized fragmentation
function, $z=E_h/(E_\mu-E_{\mu'})$ the fraction of available energy carried by
the hadron and $p_T$ the hadron transverse momentum with respect to the
virtual photon direction. $E_h$, $E_\mu$ and $E_{\mu'}$ are the energies of the
hadron, the muon before and after the scattering, respectively. As can be seen from equation~(\ref{equ:Collins}), the
Collins asymmetry appears as a $\sin\Phi_{Coll}$ modulation in the number of produced
hadrons. By measuring the Collins asymmetry on a proton and a deuteron target, the contributions from $u$- and
$d$-quarks can be disentangled \cite{Bacchetta}.

The hadron sample on which the single hadron asymmetries are computed consists
of all charged hadrons originating from the reaction vertex with $p_T>0.1$~GeV/c 
and $z>0.2$.  The Collins asymmetry is evaluated as a function of
$x$, $z$, and $p_T$ integrating over the other two variables.  The extraction
of the amplitudes is then performed fitting the expression for the transverse
polarization dependent part of the semi-inclusive DIS cross section \cite{Boer}
to the measured count rates in the target cells by a unbinned extended maximum
likelihood fit, taking into account the spectrometer acceptance. The results
have been checked by several other methods described in Ref.~\cite{COMPASS, COMPASS2}.

In the upper panel of Fig.\ref{Collins} the results for the Collins asymmetry on a proton target
are shown as a function of $x$, $z$, and $p_T$ for positive and negative 
hadrons. For small $x$ up to $x=0.05$ the measured asymmetry is small and
statistically compatible with zero, while in the last points an asymmetry
different from zero is visible. The asymmetry increases up to about $8$\% with
opposite sign for negative and positive hadrons. This result confirms the
measurement of a sizable Collins function and transversity distribution. The
asymmetry measured on the deuteron target shown in the lower panel of
Fig.\ref{Collins} is small. From this measurement the opposite sign of $u$- and $d$- quark 
transversity has been derived \cite{COMPASS}.
Both datasets have been employed in global
fits taking into account the Collins fragmentation function from BELLE and the
Collins asymmetries from COMPASS and HERMES to obtain constrains to the
transversity distribution for $u$- and $d$-quarks \cite{Bacchetta}.

\section{Two-hadron asymmetry}

The chiral-odd transversity distribution $\Delta_T q(x)$ can also be measured
in combination with the chiral-odd polarized two-hadron interference fragmentation 
function $H^{\sphericalangle}_1 (z,M^2_{inv})$ in SIDIS. $M_{inv}$ is the invariant mass of the
$h^+h^-$ pair. 
The fragmentation of a transversely polarized quark into two unpolarized
hadrons leads to an azimuthal modulation in $\Phi_{RS} = \phi_R + \phi_S -
\pi$ in the SIDIS cross section. 
Here $\phi_R$ is the azimuthal angle between $\vec R_T$ and the lepton scattering plane and 
$\vec R_T$ is the transverse component of $\vec R$ defined as:
\begin{equation}
\vec R = (z_2\cdot \vec p_1 - z_1 \cdot \vec p_2)/(z_1+z_2).
\end{equation}
 $\vec p_1$ and $\vec p_2$ are the momenta in the laboratory frame of $h^+$
and $h^-$ respectively. This definition of $\vec R_T$ is invariant
under boosts along the virtual photon direction.

The number of produced oppositely charged hadron pairs $N_{h^+h^-}$ can be written as:
\begin{equation}
N_{h^+h^-} =N_0 \cdot ( 1 + f \cdot P_t \cdot D_{NN} \cdot A_{RS} \cdot \sin \Phi_{RS} \cdot \sin
\theta).
\end{equation}
Here, $\theta$ is the angle between the momentum vector of $h^+$ in
the center of mass frame of the $h^+h^-$-pair and the momentum vector of
the two hadron system \cite{Bacchetta}.

\begin{figure}\centerline{
    \hspace*{-1cm} \includegraphics[width=1.05\columnwidth]{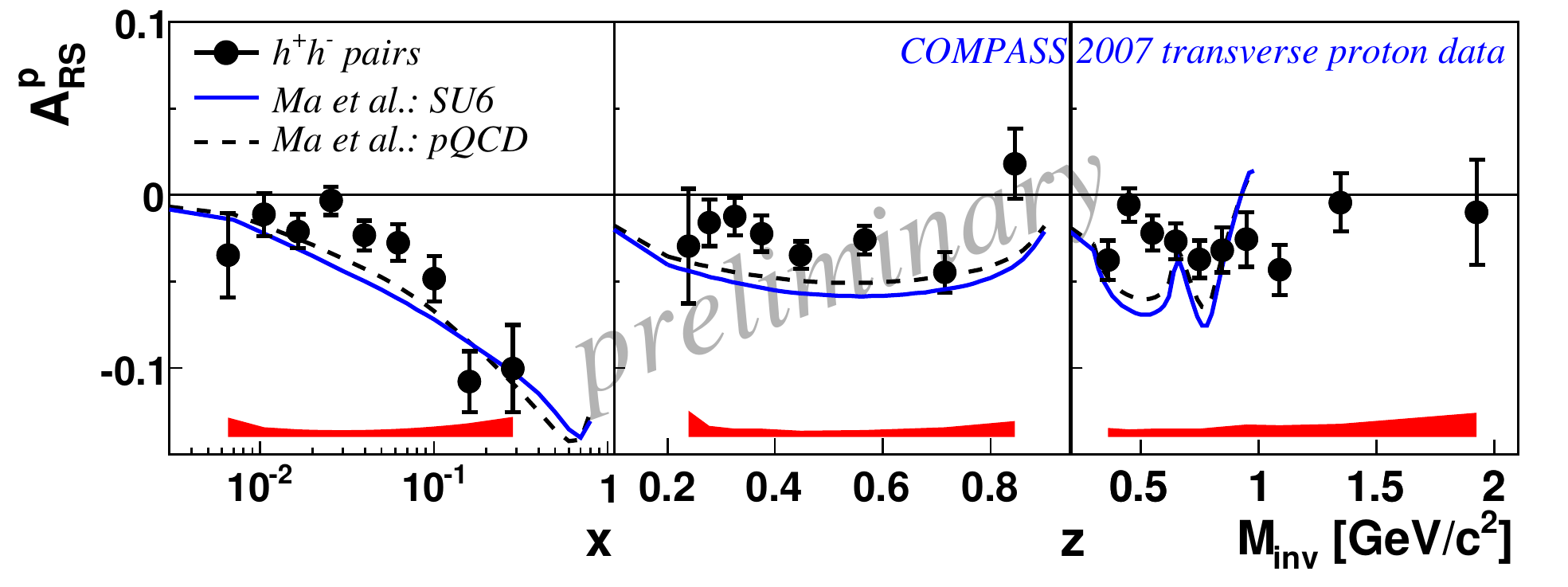}}
    \centerline{
     \includegraphics[width=\columnwidth]{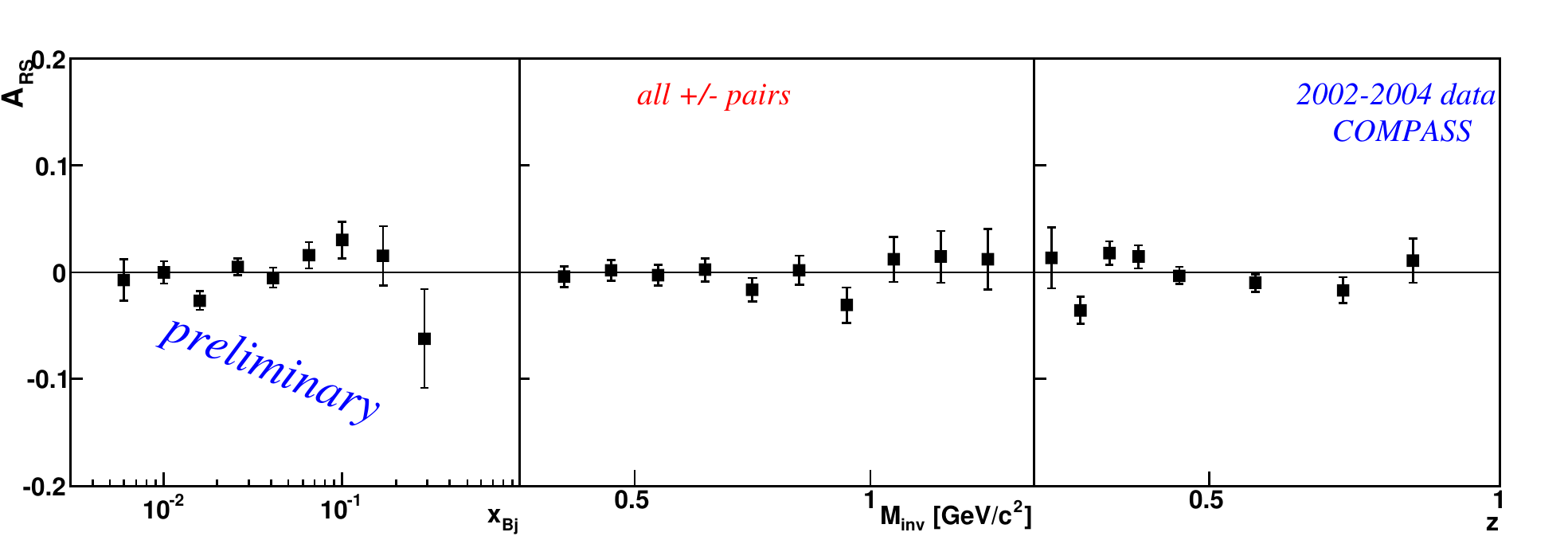}}
     \caption{{\it Upper panel:} Two-hadron asymmetry $A_{RS}$ on the proton as a function of $x$, $z$ and $M_{inv}$, compared to 
     predictions of \cite{Ma}. The lower bands indicate the
     systematic uncertainty of the measurement.
     {\it Lower panel:} Two-hadron asymmetry $A_{RS}$ on the deuteron as a function of $x$, $M_{inv}$  and $z$.
     }
	\label{pic:results}
\end{figure}

The measured amplitude $A_{RS}$ is proportional to the product of the
transversity distribution and the polarized two-hadron interference fragmentation function 
\begin{equation}
A_{RS} \propto \frac {\sum_q e_q^2 \cdot \Delta_T q(x) \cdot H^{\sphericalangle}_1(z,M^2_{inv})}
 {\sum_q e_q^2 \cdot q(x) \cdot D_q^{2h}(z,M^2_{inv})}.
\end{equation}
$D_q^{2h}(z,M^2_{inv})$ is the unpolarized two-hadron interference fragmentation function.
The polarized two-hadron interference fragmentation function
$H^{\sphericalangle}_1$ can be expanded in the relative
partial waves of the hadron pair system, which up to the
p-wave level gives~\cite{Bacchetta}:
\begin{equation}
H^{\sphericalangle}_1 = H^{\sphericalangle,sp}_1 + \cos  \theta \cdot H^{\sphericalangle,pp}_1.
\end{equation}
Where $H^{\sphericalangle,sp}_1$ is given by
the interference of $s$ and $p$ waves, whereas the function
$H^{\sphericalangle,pp}_1$ originates from the interference of two $p$
waves with different polarization. For this analysis the results are
obtained by integrating over $\theta$. The $\sin \theta$
distribution  is strongly peaked at one
and the $\cos \theta$ distribution is symmetric around zero.

Both the interference fragmentation function $H_1^\sphericalangle(z,M_{inv}^2)$
and the corresponding spin averaged fragmentation function into two hadrons
$D_q^{2h}(z, M_{inv}^2)$ are unknown. The  interference fragmentation function
$H_1^\sphericalangle(z,M_{inv}^2)$  can be measured in $e^+e^-$ annihilation or
needs to be evaluated using models \cite{Bacchetta,Jaffe,Bianconi,Radici,Metz}.

For data selection, the hadron pair sample consists of all oppositely charged
hadron pair combinations originating from the reaction vertex. The
hadrons used in the analysis have $z > 0.1$ and $x_F > 0.1$. Both cuts
ensure that the hadron is not produced in the target
fragmentation. To reject exclusively produced $\rho^0$-mesons, a cut on
the sum of the energy fractions of both hadrons was applied
$z_1+z_2<0.9$. Finally, in order to have a good definition of the
azimuthal angle $\phi_R$ a cut on $R_T > 0.07$\,GeV/c was applied.

The two-hadron asymmetry on the proton as a function of $x$, $z$ and $M_{inv}$
is shown in the upper panel of Fig.~\ref{pic:results}. A strong asymmetry in
the valence $x$-region is observed, which implies a non-zero transversity
distribution and a non-zero polarized two hadron interference fragmentation
function  $H^{\sphericalangle}_1$. In the invariant mass binning one observes a
strong signal around the $\rho^0$-mass and the asymmetry is negative over the
whole mass range. The lines are calculations from Ma {\it et al.}, based on a
SU6 and a pQCD model for transversity \cite{Ma}. The calculations can describe
the magnitude and the $x$-dependence of the measured asymmetry, while there are
discrepancies in the $M_{inv}$-behavior. The two-hadron asymmetry on the
deuteron is shown in the lower panel of Fig.~\ref{pic:results}. The measured
asymmetry is small, which shows an opposite sign of the $u$- and
$d$- quark transversity distribution.

\section{Transverse $\Lambda$ and $\bar\Lambda$ polarization}

\begin{figure}   \centerline{  
\includegraphics[width=0.9\textwidth]{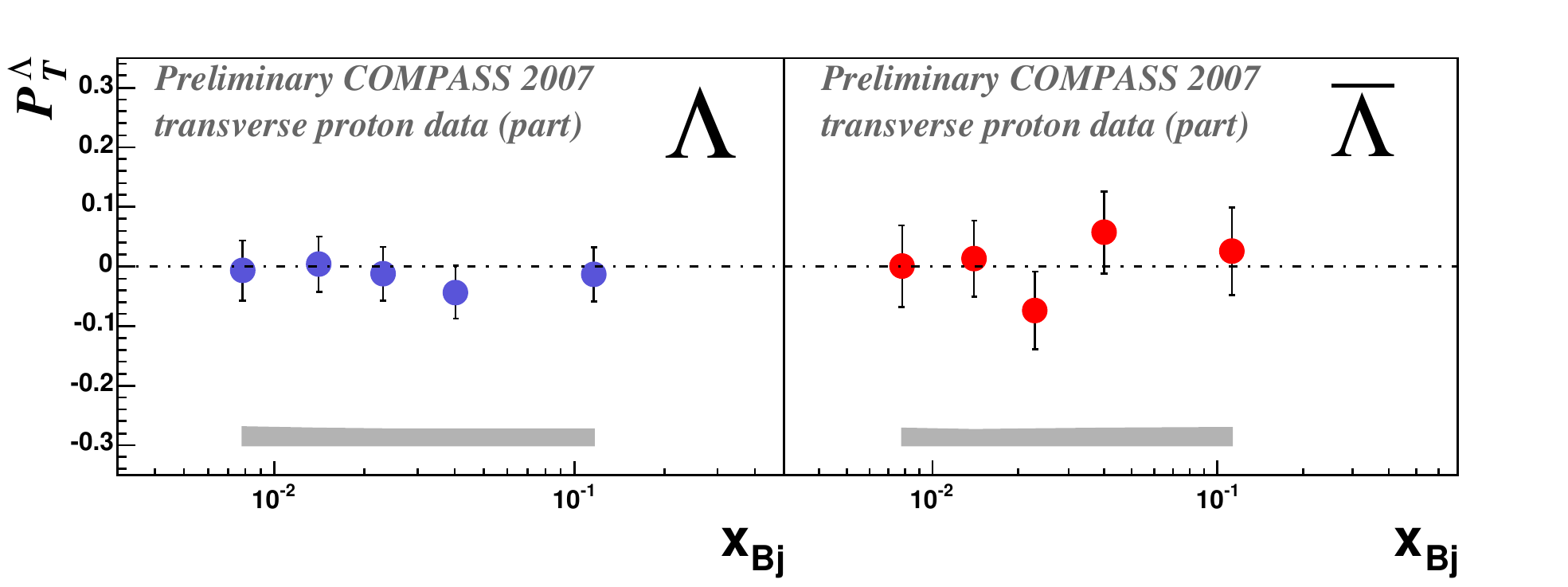}}\centerline{
\includegraphics[width=0.9\textwidth]{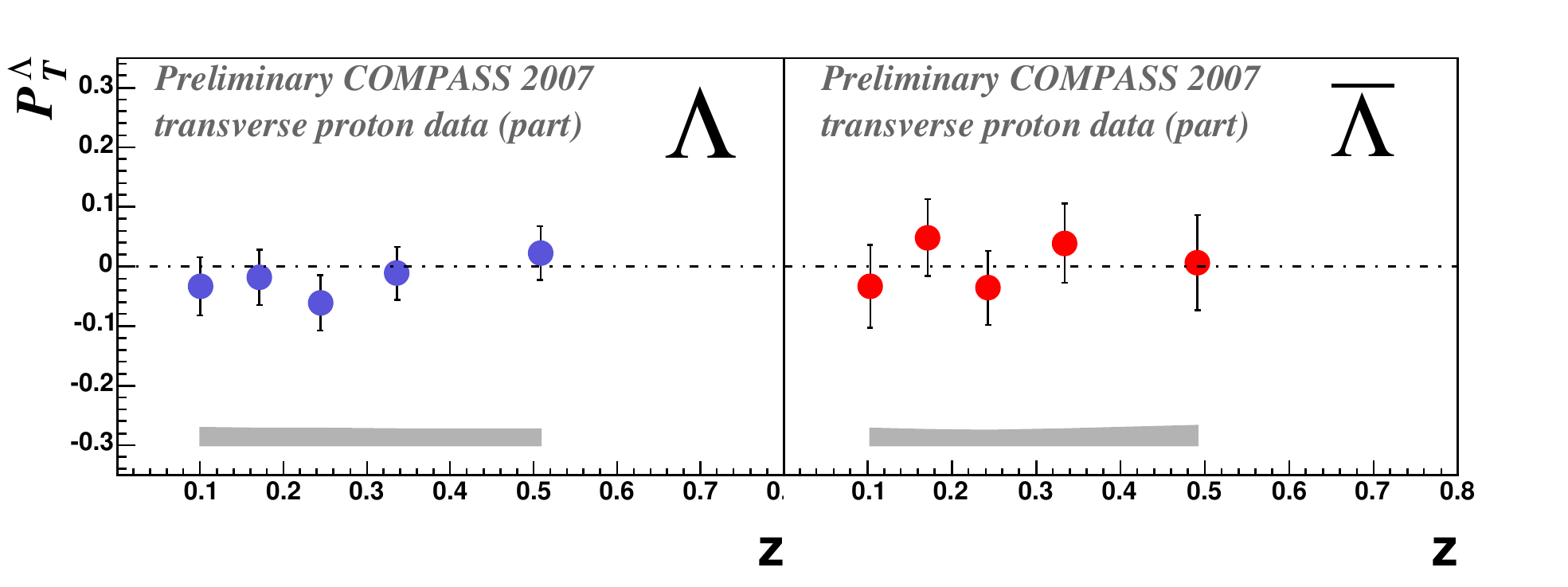}}
\caption{Transverse $\Lambda$ and $\bar\Lambda$ polarizations scattering off a transversely polarized
proton target as a function of $x$ (upper panel) and $z$ (lower panel).}
\label{Lambda}
\end{figure}

Information on the transversely polarized quark distributions $\Delta_Tq(x)$ in
the nucleon can be accessed by measuring the transverse $\Lambda$ and
$\bar\Lambda$ polarizations, which refer to a spin correlation between the
transversely polarized nucleon and the $\Lambda$ particle. $\Lambda$ and 
$\bar\Lambda$ particles are identified from their weak decays $\Lambda\rightarrow p\pi^-$
and  $\bar\Lambda\rightarrow \bar{p}\pi^+$ The $\Lambda$ polarization is
measured with respect to the reference axis $T$, where $T$ is the 
transverse quark polarization vector after 
the scattering with respect to the $\mu-\mu`$ scattering plane. 
$P^\Lambda$ is accessible through the angular distribution of
the parity violating weak decay in the $\Lambda$ rest frame by 
\begin{equation}
\frac{dN}{d \cos\theta}=\frac{N}{2}(1\pm \alpha P^\Lambda \cos \theta),
\end{equation} 
where $N$ is the number of produced $\Lambda(\bar\Lambda)$
hyperons, $\theta$ is the decay angle of the proton with respect to the
reference axis $T$. $\alpha=0.642\pm0.013$ is the analyzing power of the
parity violating $\Lambda$ decay. To determine the number of
$\Lambda$ hyperons in each $\cos\theta$ bin, a side-bin subtraction method is
used. The data of the target cells with different polarizations and the data
taking periods in which the polarization in the cells have been reversed are
used in order to cancel acceptance effects and leave only the counting rate
asymmetry $\epsilon_T(\theta)=\alpha P_T^\Lambda \cos \theta$. Finally, the
transverse $\Lambda$ and $\bar\Lambda$ polarization is extracted from the slope
of the  $\epsilon_T(\theta)$ distribution.

In analysis,  $\Lambda$s and $\bar \Lambda$s are identified from their
weak decays $\Lambda\rightarrow p\pi^-$ and  $\bar\Lambda\rightarrow
\bar{p}\pi^+$. Only tracks with momenta greater than $1$~GeV/c are selected, to
provide good tracking efficiency. A collinearity angle between the lambda
direction calculated by the position of the primary and secondary vertex and
the direction of the reconstructed lambda from its decay products has to be
within $10$~mrad. The contamination from $e^+e^-$ pairs from photon conversion
is reduced by requiring a minimal transverse momentum $p_T>23$~MeV/c of hadrons
with respect to the reconstructed $\Lambda (\bar\Lambda)$ momentum. Information
from the RICH detector is used to apply a veto condition on the proton from the
hyperon decay to reject electrons, pions and kaons.

The transverse $\Lambda$ and $\bar\Lambda$ polarization  as a function of $x$
and $z$ is shown in Fig.~\ref{Lambda}. Both the $\Lambda$ and the $\bar\Lambda$ polarization show no
significant deviation from zero in the whole explored kinematic range $8\cdot
10^{-3}<x<0.1$. The measured zero value for $P_T^\Lambda$ might be due to the
smallness of the transversity distribution in the available $x$-range or to the
fact that the polarized fragmentation function into a transverse lambda are
small in the COMPASS kinematic range, since the measured asymmetry is
proportional to the product of $\Delta_Tq(x)$ and $\Delta_TD_q^\Lambda(z)$. In
order to gain further information from the measurement of transverse $\Lambda
(\bar\Lambda)$ polarization, the limited $\Lambda$ statistics in the valence quark
region, where transversity is sizable, needs  to increased. The new data set
from 2010 with a transversely polarized proton target will contribute to that.

\section{The Sivers asymmetry} 

\begin{figure}
\includegraphics[width=\textwidth]{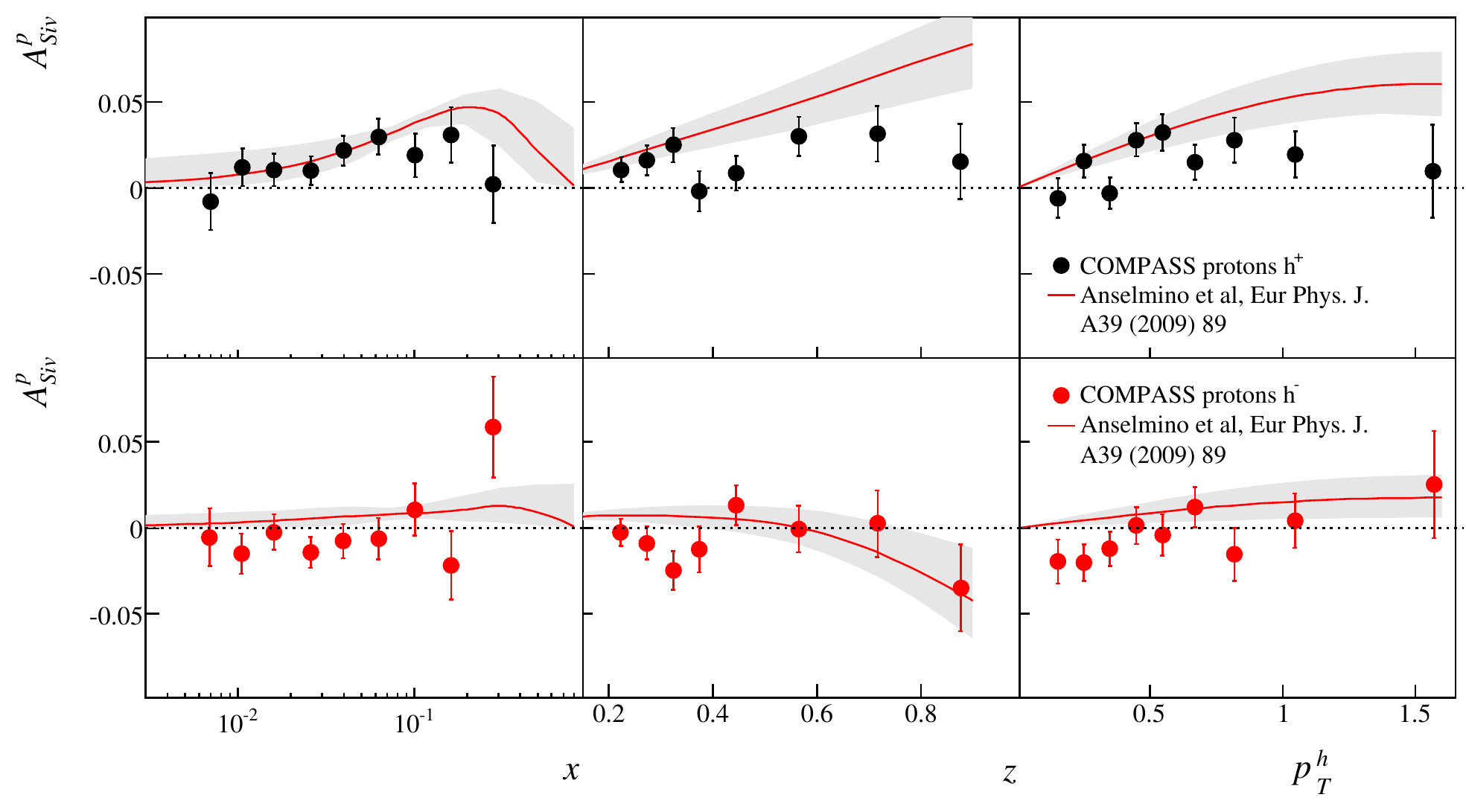}
\includegraphics[width=\textwidth]{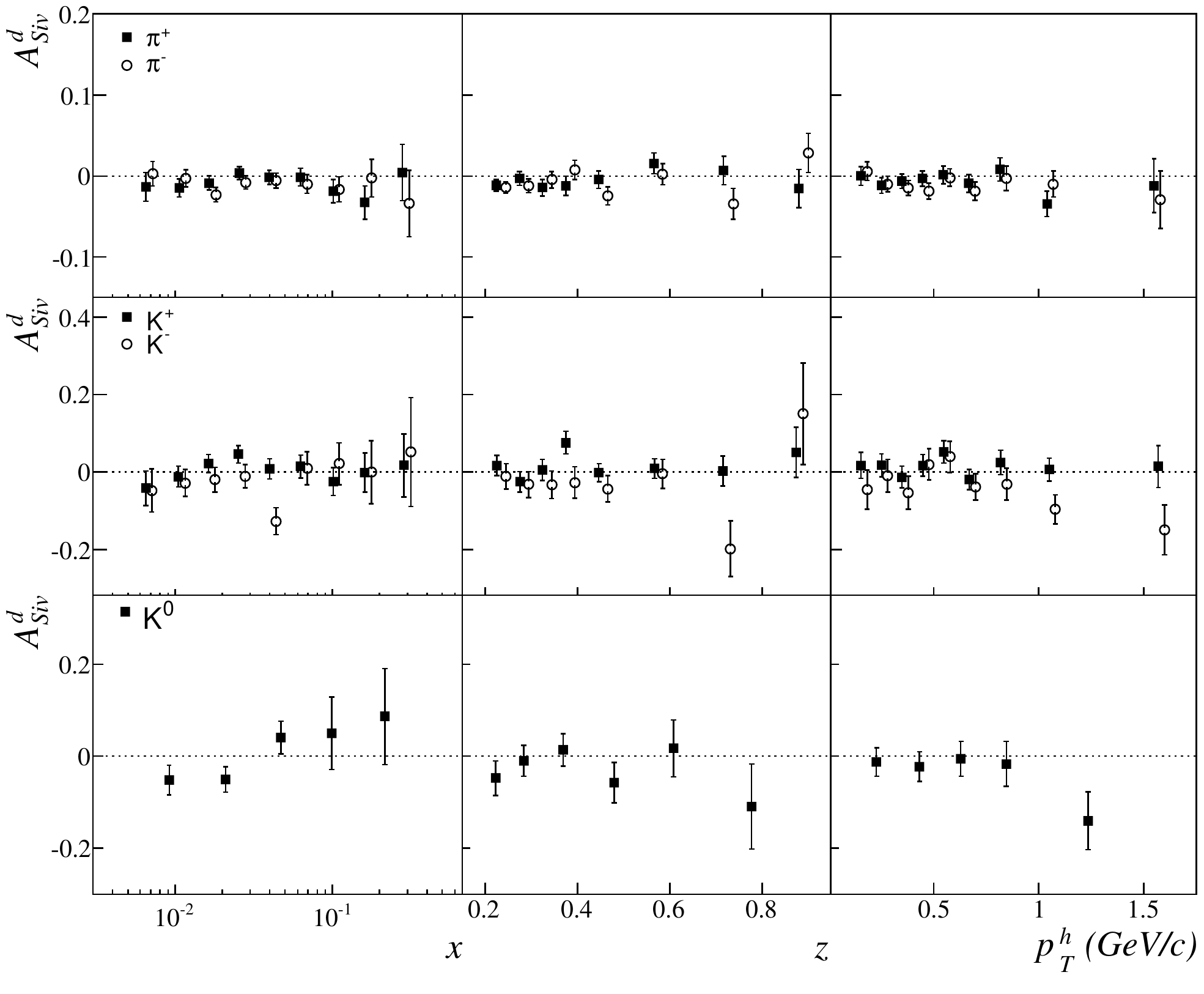}
\caption{{\it Upper panel:} Sivers asymmetry on the proton for unidentified positive (black) and negative
(red) hadrons as a function of $x$, $z$, and $p_T$ as published in \cite{COMPASS2}. Predictions from \cite{AnselminoS}
are drawn as curves. {\it Lower Panel:} Sivers asymmetry on the deuteron for
positive (filled squares) and negative (open circles) pions and kaons and 
$K^0$ as a function of $x$, $z$, and $p_T$ as published in \cite{COMPASS}. }
\label{Sivers}
\end{figure}

Another source of azimuthal asymmetry is related to the Sivers effect. The
Sivers asymmetry rises from a coupling of the intrinsic transverse
momentum $\overrightarrow{k}_T$ of unpolarized quarks with the spin of a
transversely polarized nucleon \cite{Sivers}. The correlation between the transverse nucleon
spin and the transverse quark momentum is described by the Sivers distribution
function  $\Delta_0^Tq(x, \overrightarrow{k}_T)$. The Sivers effect results in an
azimuthal modulation of the produced hadron yield:
\begin{equation}
N(\Phi_{Siv})=N_0\cdot (1+f\cdot P_t\cdot A_{Siv}\cdot \sin \Phi_{Siv}).
\label{equ:Sivers}
\end{equation}
The Sivers angle is defined as $\Phi_{Siv}=\phi_h-\phi_S$. The measured Sivers
asymmetry $A_{Siv}$ can be factorized into a product of the Sivers distribution function and the
unpolarized fragmentation function $D_q^h(z)$:
\begin{equation}
A_{Siv}=\frac{\sum_q\, e_q^2\cdot \Delta_0^Tq(x, \overrightarrow{k}_T)\cdot  D_q^h(z)}
{\sum_q \,e_q^2\cdot  q(x)\cdot D_q^h(z)}.
\end{equation}
In this case the asymmetry $A_{Siv}$ shows up as the amplitude of a $\sin\Phi_{Siv}$ modulation in the
number of produced hadrons. 

Since the Collins and Sivers asymmetries are independent azimuthal modulations
of the cross section for semi-inclusive deep-inelastic scattering \cite{Boer},
both asymmetries are determined experimentally in a common fit to the
same dataset, taking into account the acceptance of the spectrometer.

In the upper panel of Fig.\ref{Sivers} the results for the Sivers asymmetry on the proton are shown as a
function of $x$, $z$, and $p_T$.  The Sivers asymmetry for negative hadrons 
is small and
statistically compatible with zero. For  positive hadrons the Sivers asymmetry
is positive. Predictions of the Sivers asymmetry for COMPASS
kinematics from \cite{AnselminoS} are shown as curves. The predictions are
obtained using the COMPASS results for the Sivers asymmetry on the deuteron
target \cite{COMPASS} and the HERMES results on a proton target \cite{HERMES}.
The predictions describe well the x-dependence of the measures asymmetries,
while there are discrepancies in the $z$ and $p_T$ behavior. The Sivers asymmetry on the deuteron target is
shown in the lower panel of Fig.\ref{Sivers}. The asymmetry is small, which shows the opposite sign of the $u-$ and $d-$quark
 Sivers function.

\section{Azimuthal asymmetries in SIDIS off an un\-pola\-rized target}

The cross-section for hadron production in lepton-nucleon SIDIS $\ell N
\rightarrow \ell' h X$ for unpolarized targets and an unpolarized or
longitudinally polarized beam has the following form~\cite{Bacchetta2}:

\begin{equation}
\begin{array}{lcr}\displaystyle
\frac{d\sigma}{dx dy dz d\phi_h dp^2_{h,T}} =
  \frac{\alpha^2}{xyQ^2}
\frac{1+(1-y)^2}{2} \cdot\\[2ex] \displaystyle [ F_{UU,T} + 
  \varepsilon F_{UU,L} + \varepsilon_1 \cos \phi_h F^{\cos \phi_h}_{UU} \\[2ex]
   + \varepsilon_2 \cos(2\phi_h) F^{\cos\; 2\phi_h}_{UU}
   + \lambda_\mu
  \varepsilon_3
  \sin \phi_h F^{\sin \phi_h}_{LU} ]
\end{array}
\end{equation}
where $\alpha$ is the fine structure constant. 
$F_{UU,T}$,  $F_{UU,L}$, $F^{\cos \phi_h}_{UU}$,  $F^{\cos\;
  2\phi_h}_{UU}$ and $F^{\sin \phi_h}_{LU}$ are structure functions. Their 
first and second subscripts indicate the beam and target polarization,
respectively, and the last subscript denotes, if present, the
polarization of the virtual photon.  $\lambda_\mu$ is the 
longitudinal beam polarization and: 
\begin{equation}
\begin{array}{rcl}
\varepsilon_1 & = & \displaystyle\frac{2(2-y)\sqrt{1-y}}{1+(1-y)^2} \\[2ex]
\varepsilon_2 & = & \displaystyle\frac{2(1-y)}{1+(1-y)^2} \\[2ex]
\varepsilon_3 & = & \displaystyle\frac{2 y \sqrt{1-y}}{1+(1-y)^2}
\end{array}
\end{equation}
are depolarization factors. 

The Boer-Mulders parton distribution function contributes to  both the $\cos \phi_h$
and the $\cos 2\phi_h$ moments. Another source of  $\cos \phi_h$ and
the $\cos 2\phi_h$ moments in unpolarized scattering is the so-called  Cahn
effect~\cite{Cahn} which arises from the fact that the kinematics is non
collinear when the transverse momentum $k_\perp$  of the quarks is taken into
account. Additionally, perturbative gluon radiation, resulting in higher order
$\alpha_s$ QCD processes, contributes to the observed  $\cos \phi_h$ and the
$\cos 2\phi_h$ moments as well. pQCD effects become important for high
transverse momenta $p_T$ of the produced hadrons.

In analysis, data taken with a longitudinally or transversely polarized
$^6$LiD target in the year $2004$ has been spin-averaged in order to obtain
an unpolarized data sample. A Monte Carlo simulation is used
to correct for acceptance effects of the detector. The SIDIS event generation
is performed by  the LEPTO generator~\cite{lepto}, the experimental setup and the
particle interactions in the  detectors are simulated by the COMPASS Montecarlo
simulation program COMGEANT. 

The  acceptance of the detector as a function of the azimuthal angle $A(\phi_h)$ is then calculated as the
ratio of  reconstructed over generated events for each bin of $x$, $z$ and $p_T$ in which the
asymmetries are measured. The measured distribution, corrected for acceptance, is fitted with the
following functional form:
\[
\begin{array}{lll}
N(\phi_h) &=&N_0 \left( 1 + A^D_{\cos \phi}  \cos \phi_h + \right.  \\&& \left. A^D_{\cos 2\phi}
\cos 2\phi_h
 +  A^D_{\sin \phi} \sin \phi_h \right) 
\end{array}
\]
The contribution of the acceptance corrections to the systematic error was 
studied in detail.

The $\sin \phi_h$ asymmetries measured by COMPASS, not shown here,  are compatible
with zero, at the present level of statistical and systematic errors, over the
full range of $x$, $z$ and $p_T$ covered by the data.

\begin{figure}
\begin{center}
\includegraphics[width=0.9\columnwidth]{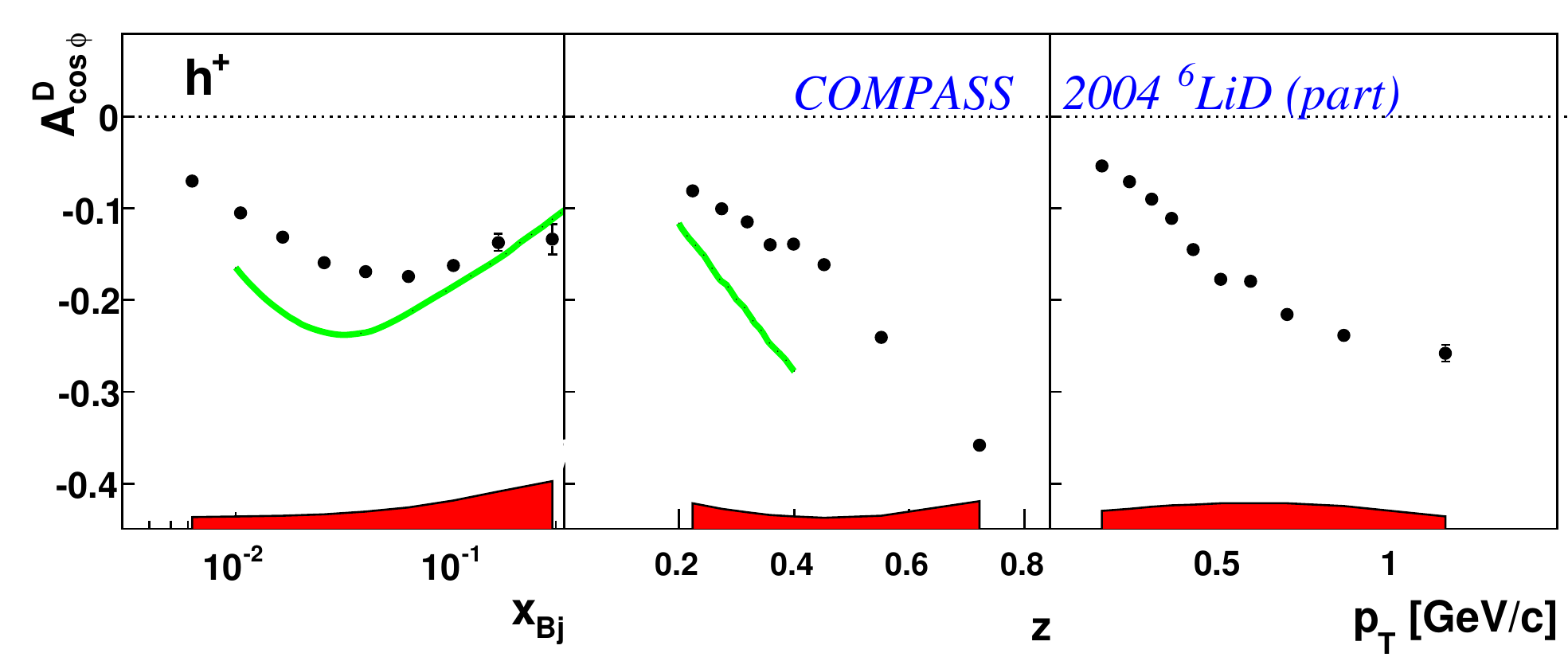}
\includegraphics[width=0.9\columnwidth]{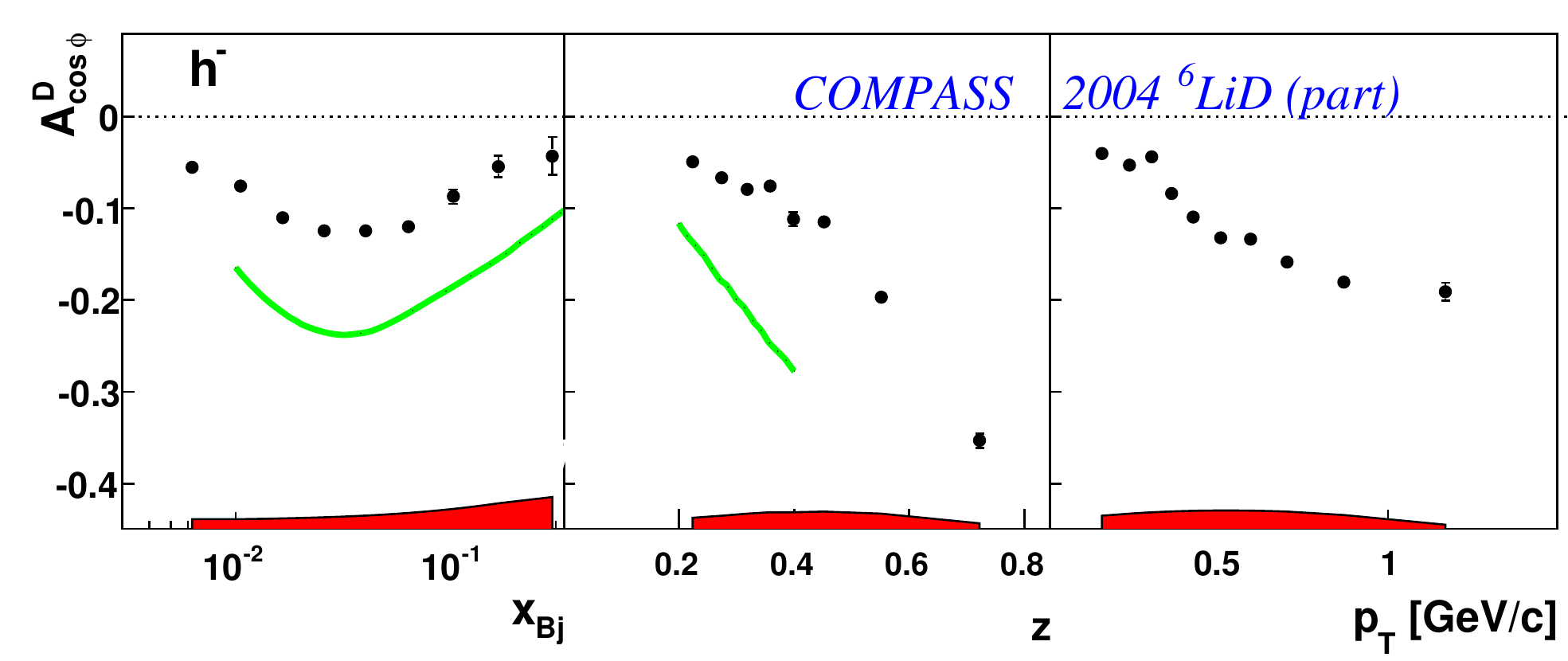}
\end{center}
\caption{$\cos \phi_h$ asymmetries from COMPASS deuteron data
for positive (upper row) and negative (lower
row) hadrons; the asymmetries are divided by the kinematic factor
$\varepsilon_1$ and the bands indicate the size of the systematic uncertainty. 
The superimposed curves are the values predicted by~\protect\cite{anselmino2}
taking into account the Cahn effect only.
}
\label{f:cosphi}
\end{figure}

The $\cos \phi_h$ asymmetries extracted from COMPASS deuteron data
are shown in Fig.~\ref{f:cosphi} for positive (upper row) and negative (lower
row) hadrons, as a function of $x$, $z$ and $p_T$. The bands indicate the size
of the systematic error. The asymmetries show the same trend for positive and
negative hadrons with  slightly larger absolute values for  positive hadrons. 
Values as large as 30$-$40\% are reached in the last point of the $z$ range. 
The theoretical
prediction~\cite{anselmino2} in Fig.~\ref{f:cosphi} takes into account the Cahn
effect only, which
does not depend on the hadron charge. The Boer-Mulders parton distribution function 
is not
considered in this prediction.

\begin{figure}\begin{center}
\hspace*{1cm}\includegraphics[width=\textwidth]{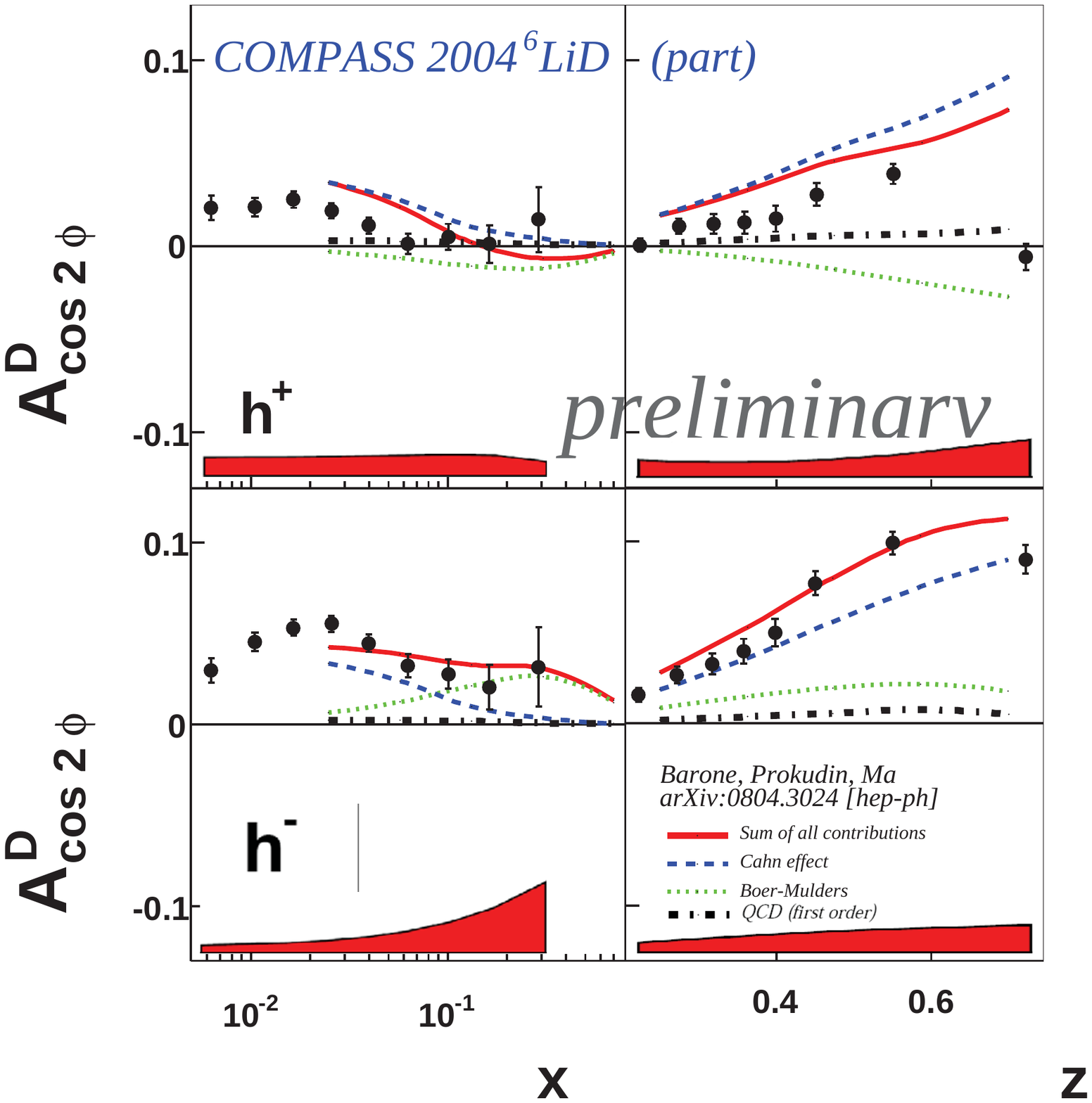}
\end{center}
\caption{$\cos 2 \phi_h$ asymmetries from COMPASS deuteron data
for positive (upper row) and negative (lower
row) hadrons; the bands indicate size of the systematic error. }
\label{f:cos2phi}
\end{figure}

The $\cos 2 \phi_h$ asymmetries are shown in Fig.~\ref{f:cos2phi} together with the
theoretical predictions of~\cite{barone}, which take into account the kinematic
contribution given by the Cahn effect, first order pQCD (which, as expected, is
negligible in the low $p_T$ region), and  the Boer-Mulders parton distribution
function (coupled to the Collins fragmentation function), which gives a different
contribution to positive and negative  hadrons. 

In~\cite{barone}, the Boer-Mulders
parton distribution function is assumed to be proportional to the Sivers function as
extracted from preliminary HERMES data. The COMPASS data show an  amplitude
different  for positive and negative hadrons, a trend which confirms the theoretical
predictions. There is a satisfactory agreement between the data points and the model
calculations, which hints to a non-zero Boer-Mulders parton distribution function.

\section{Summary and Outlook}


Results for the Collins  and the two-hadron azimuthal asymmetry at COMPASS in
semi-inclusive deep-inelastic scattering off transversely polarized proton and deuteron 
targets have been presented. 
For $x>0.05$, a Collins and a two-hadron asymmetry
different from zero and increasing magnitude with increasing $x$-Bjorken have
been observed on the proton. The asymmetries on the deuteron are small and compatible with zero.
The transverse $\Lambda$ and the $\bar\Lambda$ polarization on the proton were
found to be small and compatible with zero within the available kinematic
range.

The measured Sivers asymmetry on the proton for negative hadrons is compatible
with zero, while a positive asymmetry is observed for positive hadrons. 
On the deuteron, the Sivers asymmetry is small. 

The measured unpolarized azimuthal asymmetries on a deuteron target show large
$\cos\phi_h$ and $\cos 2\phi_h$ moments which can be qualitatively described in
model calculations taking into account the Cahn effect and the intrinsic $k_T$
of the quarks in the nucleon and the Boer-Mulders structure function.

With the data from a full-year transverse-target running completed in $2010$,
COMPASS will significantly increase its statistical precision in all
measurements of transverse-spin dependent asymmetries.

\section*{Acknowledgments}

This work has been supported in part by the German BMBF.


\end{document}